\documentclass[twocolumn]{aa} 

\def\specchar#1{{\sc #1}}

\def\arcsec{\hbox{$^{\prime\prime}$}}
\def\FeI{\mbox{Fe\,\specchar{i}}}
\def\NiI{\mbox{Ni\,\specchar{i}}}
\def\NaI{\mbox{Na\,\specchar{i}}}
\def\KI{\mbox{K\,\specchar{i}}}

\def\BaII{\mbox{Ba\,\specchar{ii}}}
\def\CaIIK{\mbox{Ca\,\specchar{ii}\,\,K}}       

\def\ms{\hbox{m$\;$s$^{-1}$}}

\usepackage{natbib}
\usepackage{graphicx}
\usepackage{color}

\begin{document}
   \title{Properties of convective motions in facular regions}

\author{R. Kostik\inst{1}, E. Khomenko\inst{2,3,1}}

\institute{Main Astronomical Observatory, NAS, 03680, Kyiv,
Ukraine\\ \email{kostik@mao.kiev.ua} \and Instituto de
Astrof\'{\i}sica de Canarias, 38205 La Laguna, Tenerife, Spain
\and Departamento de Astrof\'{\i}sica, Universidad de
La Laguna, 38205, La Laguna, Tenerife, Spain\\
}

\date{Received XXX; accepted xxx}

\abstract
{}
{In this paper, we study the properties of solar granulation in a facular
region from the photosphere up to the lower chromosphere. Our aim is to
investigate the dependence of granular structure on magnetic field
strength.}
{We use observations obtained at the German Vacuum Tower Telescope
(Observatorio del Teide, Tenerife) using two different instruments: Triple
Etalon SOlar Spectrometer (TESOS), in the \BaII\ 4554 \AA\ line to measure
velocity and intensity variations along the photosphere; and, simultaneously,
Tenerife Infrared Polarimeter (TIP-II), in the \FeI\ 1.56 $\mu$m lines to the
measure Stokes parameters and the magnetic field strength at the lower
photosphere.}
{We obtain that the convective velocities of granules in the facular area
decrease with magnetic field while the convective velocities of intergranular
lanes increase with the field strength. Similar to the quiet areas, there is
a contrast and velocity sign reversal taking place in the middle photosphere.
The reversal heights depend on the magnetic field strength and are, on
average, about 100 km higher than in the quiet regions. The correlation
between convective velocity and intensity decreases with magnetic field at
the bottom photosphere, but increases in the upper photosphere. The contrast
of intergranular lanes observed close to the disc center is almost
independent of the magnetic field strength. }
{The strong magnetic field of facular area seems to stabilize the convection
and to promote more effective energy transfer in the upper layers of the
solar atmosphere, since the convective elements reach larger heights.}


\maketitle

\section{Introduction}

Solar faculae are areas on the solar surface surrounding active regions and
appearing bright toward the limb. Understanding their properties is important
from several points of view. On the one hand, facular contrast influences
total solar irradiance variations. On the other hand, faculae are believed to
be composed of conglomerates of magnetic elements. Therefore, in faculae we
can observe the interaction of convection with relatively strong magnetic
fields, which is interesting in itself from a physical point of view and also to
constrain state of art numerical models of magneto-convection.

It is well established that the origin of facular brightness excess is
related to the presence of magnetic field: the brightness excess is caused by
the modified radiative transfer through magnetic atmospheres viewed from
different angles \citep[e.g.][]{Keller+etal2004, Carlsson+etal2004,
Vogler2005, Okunev+Kneer2005, Steiner2005}. The variation of facular
brightness near the solar limb can provide information about the structure
and properties of the magnetic elements composing facular regions.
Measurements of centre-to limb variations of the contrast done by different
authors mostly agree that the contrast increases till $\mu=0.2-0.4$
\citep[][]{Muller1977, Auffret+Muller1991, Topka+etal1997, Ortiz+etal2002,
Okunev+Kneer2004, Hirzberger+Wiehr2005}. A controversy exists for the extreme
limb, where it is not clear if the contrast further increases
\citep[][]{Boer1997, Sutterlin1999} or falls off toward the limb. The variety
of results can be attributed to the difference in spatial resolution of
observations, facular size, its magnetic field strength, as well as selection
criteria. At high spatial resolution (0.\arcsec1--0.\arcsec2) and at disk
centre, facular regions appear as conglomerates of bright points, small pores
and small-scale granular structures \citep{Lites+etal2004, Berger2007,
Narayan+etal2010}. The contrast of faculae at disk centre is slightly
positive or close to zero, though some negative values are also detected
\citep[see][and references therein]{Title1992, Topka+etal1997}.

Several classes of models claim to reproduce facular brightness properties. The
``hot wall'' model, initially proposed by \citet{Spruit1976}, consists of a
magnetic flux tube, evacuated due to the presence of magnetic field, similar
to Wilson depression in sunspots. Viewed from an angle, the line-of-sight
penetrates deeper into the tube because of its lower density and, thus,
hotter layers become visible. This idea was followed in works by e.g.,
\citet{Topka+etal1997, Steiner2005, Okunev+Kneer2005}, who provide evidences
that the ``hot wall'' model can closely reproduce observed facular
properties, such as continuum contrast, its centre-to-limb variation, as well
as the dependence of contrast on the magnetic field. Much more complex models
of facular regions based on ``realistic'' MHD simulations
\citep{Keller+etal2004, Carlsson+etal2004, Vogler2005} suggest that faculae
are seen bright on the limb because hot granular walls become visible through
transparent magnetic field concentrations. The three-dimensional structure of
granules becomes apparent at faculae near the solar limb in high resolution
observations and is well reproduced in the simulations. Two factors are
crucial for the facula formation: the shape of the granule limb-ward of the
flux concentration and the size of the magnetic field concentration. If the
flux concentration is too small, the opacity reduction is not sufficient to
have the continuum intensity formed exclusively in the hot granule. A
quantitative disagreement still remains in the values of the peak brightness
in faculae, being substantially larger in simulations compared to
observations \citep{Keller+etal2004}.

\citet{Berger2007} question the dependence of granular brightness on magnetic
field strength obtained in earlier works \citep{Topka+etal1997,
Ortiz+etal2002}. \citet{Berger2007} analyze extremely high spatial resolution
data (close to 0.\arcsec1) obtained at the Swedish Solar Telescope on La
Palma. They claim that if, instead of analyzing the brightness for binned
magnetogram signals \citep[as is done in][]{Topka+etal1997}, one analyzes
magnetic flux density for facular points segmented from the data set, there
is no dependence of the brightness on the magnetic flux density.
\citet{Berger2007} propose that what we see as faculae are granular walls,
and not interiors of the magnetic flux tubes. Since granules all have similar
properties, there is no dependence of their brightness on the magnetic field,
and the magnetic field only plays an indirect role, making the atmosphere
transparent in front of the granules.

The strong magnetic field of facular regions modifies the properties of
convection. Simulations of magnetic flux emergence through the convection
zone show that during the emergence granules become larger, more elongated
and smoother \citep{Cheung+etal2008}. Observationally, in already formed
facular region granulation presents a lot of fine structuring at high spatial
resolution: isolated bright points, strings of bright points and dark
micro-pores, ribbons, or more circular flower structures \citep{Title1992,
Berger2004, Narayan+etal2010}. Granules become smaller than in quiet areas
\citep{Muller1977, Schmidt1988, Title1992} and intergranular lanes are
characterized by the presence of micro-pores. Granular velocities in facular
and plage areas are found to be generally lower than in the nearby quiet
areas \citep{Nesis+Mattig1989, Title1992, Narayan+etal2010}. Velocities
measured in plage and facular areas are found to depend on the magnetic field
strength. From medium-to-high  (430 km) resolution observations of the disk
centre plage in \NiI\ 6768 \AA\ line, \citet{Title1992} obtain an increase of
downflow velocities with magnetic field strength up to 600 G, and a decrease
for stronger fields. A similar result is reported by \citet{Montagne1996,
Berger2004, Morinaga2008, Narayan+etal2010}. The correlation between velocity
and intensity fields, typical for granulation, is found to be partially
destroyed by the magnetic field \citep{Rimmele2004, Narayan+etal2010}. One of
the possible reasons of the de-correlation, proposed by
\citet{Narayan+etal2010} is that small scale convection in plage areas does
not overshoot to the same height as field-free convection and leaves only
weak traces at the line-forming region.

It is clear that further observational studies of magneto-convection in
strongly magnetized plage and facular areas are needed to constrain the
relationship between magnetic field and granulation, as well as the
properties of magnetic elements.
Here we report on such an observational study based on state-of-the-art
simultaneous observations done at the German Vacuum Tower Telescope at the
Observatorio del Teide (Tenerife) with two instruments: TIP-II
\citep{Collados2007} and TESOS \citep{Tritschler+etal2002}.
We analyze correlation between granular/intergranular intensities,
velocities, and magnetic field; as well as the results of the heights of sign
reversals of contrast and velocity as a function of the magnetic field.

\begin{figure*}
\centering
\includegraphics[width=18cm]{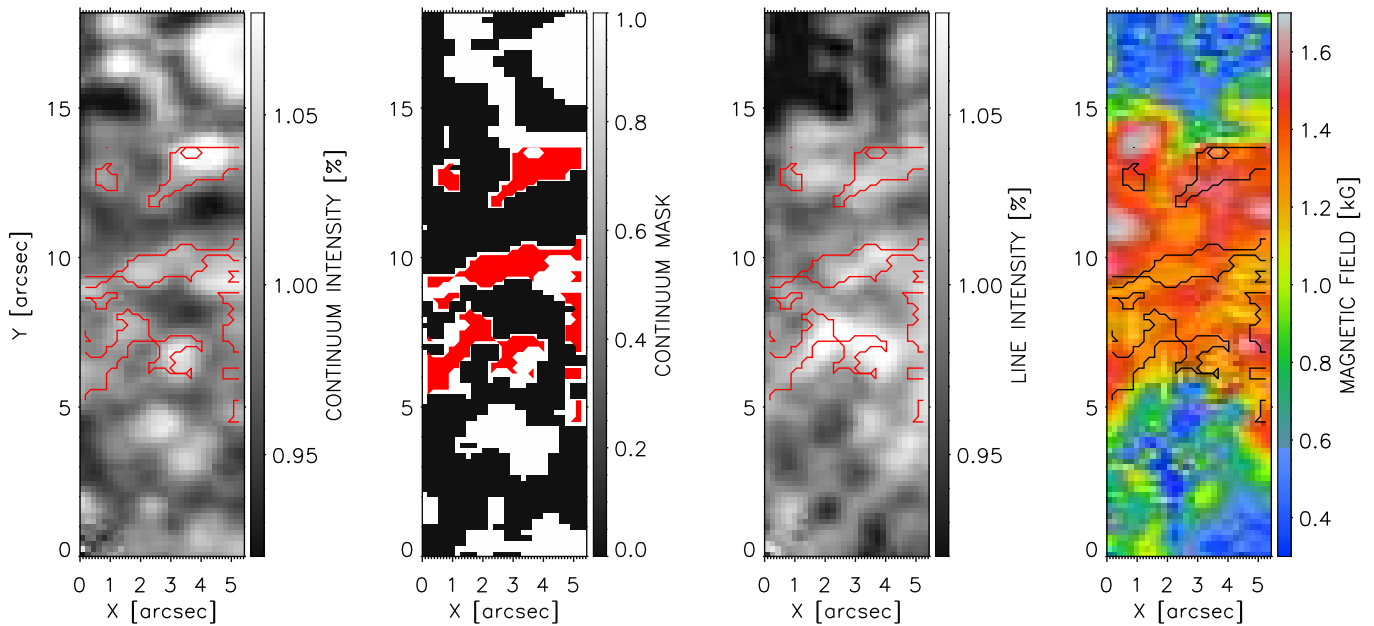}
\caption{An overview of the observations. The panels from left to right are:
\BaII\ 4554 \AA\ continuum intensity in the units of its spatially average
value; mask applied to locate granules and intergranular lanes; \BaII\ 4554
\AA\ line centre intensity in the units of its spatially average value;
magnetic field strength from the inversion of \FeI\ IR lines. Contours (same
in all panels) mark the locations of granular areas with magnetic field above
1.2 kG. } \label{fig:fov}
\end{figure*}

\section{Observations}

The observations were obtained on the 13th of November, 2007, at the German
Vacuum Tower Telescope at the Observatorio del Teide (Tenerife). We
simultaneously observed three spectral regions: \FeI\ 15643--15658 \AA,
\BaII\ 4554 \AA, and \CaIIK\ 3968 \AA.

Using the filtergrams in  \CaIIK, we have selected a facular area located
close to disk centre. There was no sunspot activity at the
time of observations and the observed facular area was not attached to any
active region.

The acquisition of the data was controlled by the Tenerife Infrared
Polarimeter \citep[TIP-II, see][]{Collados2007}, to scan repeatedly a small
area of the solar surface. The slit length was 84\arcsec\ and the slit width
0.\arcsec5. The pixel size of the CCD of TIP was 0.\arcsec185. One scan
consisted of 15 steps, separated 0.\arcsec35, with a total scanned area of
84\arcsec$\times$5.\arcsec5. The time interval between successive scan steps
was 27.3 sec. Thus, it took 6 min 50 sec to record each complete series of 15
steps. All four Stokes parameters were recorded in the 1.56 $\mu$m region. In
total, 22 scans were done. The spectral sampling of the infrared data was of
14.7 m\AA.

Simultaneously with TIP-II, the TESOS instrument was making scans through the
spectral profile of the \BaII\ 4554 \AA\ line. TESOS \citep[Triple Etalon
SOlar Spectrometer;][]{Tritschler+etal2002} is a Fabry-Perot interferometer
equipped with three etalons which allow to scan spectral lines in a
two-dimensional field of view. The CCD of TESOS had 512$\times$512 pixels of
0.\arcsec089 size. Thus, the total size of the observable area on the Sun was
45.\arcsec6$\times$45.\arcsec6. The \BaII\ 4554 \AA\ line was covered by 37
spectral points with a spectral sampling of 16 m\AA. During the scan, TIP was
moving the whole field of view of TESOS at steps of 0.\arcsec35. However, at
every moment of the observations, the area scanned by TIP and the field of
view of TESOS overlapped. Each complete scan of the 37 spectral points of the
\BaII\ line profile took 25.6 sec.

In addition to TIP and TESOS, we also recorded \CaIIK\ 3968 \AA\ filtergrams
with a pixel size of 0.\arcsec123, a temporal cadence of 4.93 seconds and a
total field of view of 110\arcsec$\times$110\arcsec.

The seeing conditions were good to moderate during the observations. The
$r_0$ parameter was around 10, indicating spatial resolution no more than
0.\arcsec5 in the blue range of the spectrum at 4500 \AA, and the resolution
around 2\arcsec\ in the infrared.
Due to jumps of the Adaptive Optics (AO) system, the observed area was lost a
few times during the observations and we could not completely recover its
position. Thus, for the further analysis we used only five strictly
co-spatial TIP scans (from 2nd to 6th), with a total duration of 34 min 41
sec.

\subsection{Data reduction}

As a first step of the reduction process, the data from all three cameras
(TIP, TESOS and \CaIIK) were corrected for the dark current and flat field,
following the standard procedure. In addition, spectropolarimetric data from
TIP were corrected for the instrumental polarization, calibrated and
de-modulated to give Stokes $I$, $Q$, $U$ and $V$ profiles, using the standard
reduction routines for this instrument.

We then performed the spatial alignment between all three data channels. Here
we should recall that TIP records all Stokes spectra simultaneously, however
each spatial slit position is separated 27.3 sec in time. In contrast, TESOS
records all spatial points simultaneously, but each spectral position is
separated 25.6/37 sec in time. The alignment of these data was far from
trivial.

Firstly, we compared the images of a reticle target, scanned during the
observations by both instruments. From them we calculated the rotation angle
and scaling factor to apply to TESOS data in order to have the same
orientation and spatial sampling as TIP. We then rotated and re-sampled TESOS
data accordingly. Secondly, we constructed ``false'' continuum Stokes $I$
maps from TIP (keeping in mind that they are not strictly simultaneous in
time), and continuum \BaII\ images from TESOS. To make a better match, we
artificially decreased the spatial resolution of the TESOS continuum maps to
make it similar to the resolution of the TIP maps in the infrared.
The TIP continuum image was ``moved'' over the TESOS continuum image to find
the location with the best correlation. This has allowed us to align both
images with a 1-pixel precision of TIP (0.\arcsec185). The common area
between both instruments had the size of 5.\arcsec5$\times$18.\arcsec5. A
similar procedure was applied to the $\CaIIK$ data, using \BaII\ line core
images as a reference to find the best correlation.

In summary, after all the reduction and alignment processes, our observational material
consists of the following datasets:
\begin{itemize}
\item Five TIP maps of Stokes spectra of \FeI\ lines at 1.56 $\mu$m,
    separated 6 min 50 sec in time and 5.\arcsec5$\times$18.\arcsec5 in
    size;
\item Time series of \BaII\ 4554 \AA\ line spectra over a
    5.\arcsec5$\times$18.\arcsec5 area, co-spatial with TIP, with a
    temporal cadence of 25.6 sec and a duration of 34 min 41 sec;
\item Time series of \CaIIK\ filtergrams over
    5.\arcsec5$\times$18.\arcsec5 area, co-spatial with TIP and TESOS,
    with a temporal cadence of 4.93 sec and the same duration as the TIP
    and TESOS series.
\end{itemize}

Figure~\ref{fig:fov} gives an example of the co-spatial maps of the continuum
and line core intensity in the \BaII\ line, the mask applied to select
granular and intergranular areas, and the magnetic field strength derived
from \FeI\ infrared (IR) spectra (see the details below). In the following we
do not use \CaIIK\ data, but we retain the complete description of our
observational dataset for a forthcoming paper. The comparison of the \BaII\
line core intensity maps and magnetic field strength map shows that areas
with larger magnetic field are, generally, brighter. The \BaII\ line core
intensity traces magnetic field concentrations in a similar way as \CaIIK\
line does.

\begin{figure*}[t]
\centering
\includegraphics[width=18cm]{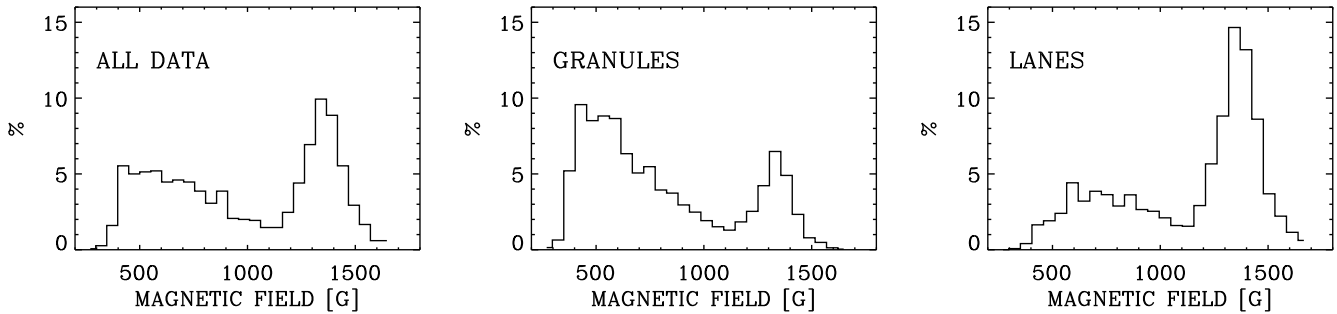}
\caption{Histograms of the magnetic field strength from the inversion of the
\FeI\ lines. Left: histogram over all spatial points; middle: only pixels
with \BaII\ continuum contrast larger than the average (granules); right:
only pixels with \BaII\ continuum contrast lower than the average
(intergranular lanes). } \label{fig:histogram}
\end{figure*}

\subsection{Calculation of magnetic field from IR \FeI\ lines}

Stokes parameters of the \FeI\ 15648 and 15652 \AA\ lines were inverted using
the SIR inversion code \citep[Stokes Inversion based on Responce functions,
see][]{RuizCobo+delToroIniesta1992}. We fitted all four Stokes parameters of
the two \FeI\ lines.
We assumed a one-component model including magnetic field. The free
parameters of the inversion were: temperature (with a maximum of 5 nodes),
line of sight and microturbulent velocities and magnetic field. The latter
was assumed vertical and constant with height. The contribution of a
non-magnetic surrounding was taken into account by varying the stray light
fraction. An average Stokes $I$ profile over the non-magnetic areas was taken
as a stray light profile. The rightmost panel of Fig.~\ref{fig:fov} gives the
magnetic field strength of the first (of the five) TIP maps used in the
current study. The observed facular region contained features with a
rather strong magnetic field.
In fact, the magnetic field strength measured directly from the splitting of
the infrared \FeI\ lines was very close to the one obtained from the
inversion. The average stray light factor reached about 30\%, and the average
flux over the area made about 300 G. The histogram of the magnetic field
strength measured in all spatial points in the five TIP scans is given in
Fig.~\ref{fig:histogram} (left panel). While there are many pixels with
intermediate field strength (300-800 G), the histogram peaks at 1300 G.

\subsection{Calculation of velocity and intensity variations from
\BaII}

The velocity and intensity variations were calculated from the \BaII\ line
profiles following the $\lambda$-meter technique \citep[see for the detailed
description][]{Stebbins+Goode1987, Shchukina+etal2009, Kostik+etal2009}. As
follows from \citet{Shchukina+etal2009}, the formation height of \BaII\ line
spans all the photosphere from 0 to almost 700 km.

We used 14 reference widths of the average \BaII\ 4554 \AA\ line intensity
profile over the whole time and space. For this average profile,
and for prescribed line widths, $W$, we calculated
the intensity value, $\bar{I}(W)$, and the displacements (velocities) in the
red and blue wing, $\bar{V_r}(W)$ and $\bar{V_b}(W)$, corresponding to each
width $W$. We repeated the same procedure for each individual \BaII\ profile
at each spatial position and time moment $t$. We then found intensity
variations $\delta I(t,x,W)$ and red and blue wing velocities $V_r(t,x,W)$,
$V_b(t,x,W)$ at the corresponding 14 reference widths. We are interested in
fluctuations of these quantities given by the following relations:
\begin{eqnarray}
\label{velo}
\delta I(t,x,W) & = & I(t,x,W) - \bar{I}(W)  \\
\nonumber
\delta V_r(t,x,W) & = & V_r(t,x,W) - \bar{V_r}(W)  \\
\nonumber
\delta V_b(t,x,W) & = & V_b(t,x,W) - \bar{V_b}(W)
\end{eqnarray}
The intensity variations are normalized to their corresponding mean value
for each $W$ level, to give the contrast variations, $\delta C$, as:
\begin{equation} \label{eq:deltac}
\delta C(t,x,W)  =  I(t,x,W)/\bar{I}(W) - 1
\end{equation}

We would like to remind here that, applying the $\lambda$-meter technique,
both intensity (contrast) and velocity fluctuate for a given fixed reference
width $W$, since this width can correspond to a higher or lower section of
the profile, and can be displaced in wavelength due to velocities.


The zero velocity reference was placed requesting $\langle \delta V_r(t,x,W)
\rangle=0$ and $\langle \delta V_b(t,x,W) \rangle=0$, averaged over time and
space for the reference width $W$ closest to the line core. This
determination of zero velocity reference is less affected by the standard
convective blue shift than a granulation-averaged intensity profile, because
to first order the granular intensity-velocity correlation does not influence
the averaged line minimum, as long as the granulation is resolved. According
to several studies \citep[e.g.][]{Lopresto+Pierce1985} the lines forming high
at the atmosphere (near the temperature minimum, as our \BaII\ line) have a
very little convective blue shift. The above cited work does not provide
information about the \BaII\ 4554 \AA\ line, however for lines formed at
similar heights (\FeI\ 3767 \AA, \NaI\ 5896 \AA, \KI\ 7699 \AA), the
remaining blueshift is around 60 \ms. Give than, we realize that our
determination of the zero velicity position may contain a blue shft, but its
value is not expected to be larger than 60 \ms.

We should keep in mind that the formation heights of the blue and red wing
intensity points of the line profile are, in general, not the same because of
the Doppler velocity shift \citep{Shchukina+etal2009}, so the red and blue
velocities may correspond to different heights. Despite this, in the rest of
the paper we use average velocities:
\begin{equation}
\delta V(t,x,W) =(\delta V_r(t,x,W) +\delta V_b(t,x,W))/2
\end{equation}

The formation heights corresponding to each width level are expected to vary
significantly in space, being different in granules and intergranular lanes.
To ascribe an approximate formation height to each of the 14 $W$ levels, we
followed NLTE (non-Local Thermodynamic Equilibrium) radiative transfer
calculations, described in previous works \citep{Shchukina+etal2009,
Kostik+etal2009}. In the following, we will refer to  ``continuum'' velocity
and ``continuum'' intensity (contrast) in \BaII\ as velocities and
intensities corresponding to the highest part of the profile, i.e., the
deepest layer in the atmosphere sampled by our procedure. Through the paper
we adopt a positive sign for upflowing velocity (motion toward the observer).

\begin{figure*}
\centering
\includegraphics[width=18cm]{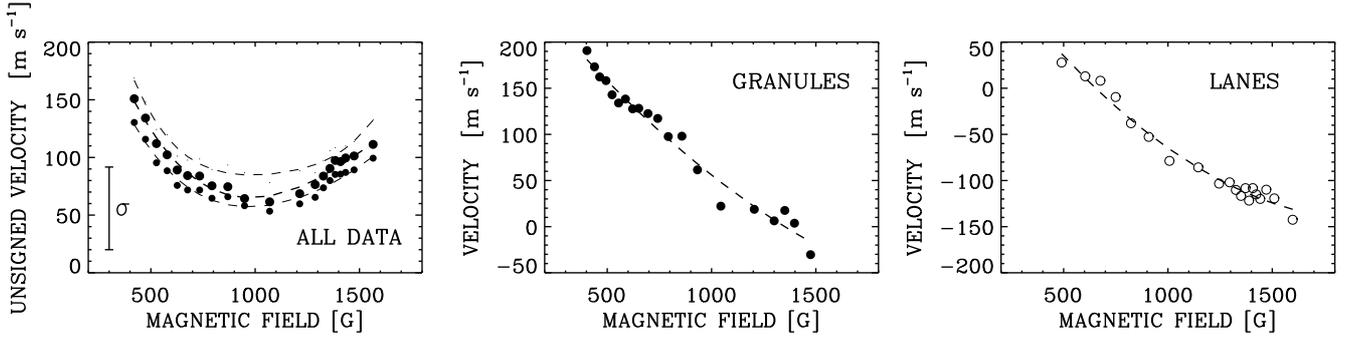}
\caption{Convective velocities measured in \BaII\ line as a function of
magnetic field strength obtained from the \FeI\ lines. Left panel: unsigned values of the
velocity, all pixels are considered. The three group of points correspond to the velocities at
three levels of the \BaII\ profile, larger points mean higher in the
atmosphere (0, 400 and 650 km, approximately). Middle: only granular points,
velocities at the deepest level of the atmosphere. Right panel: only
intergranular points. } \label{fig:velocity}
\end{figure*}

\subsection{Separation of convective and wave components}

Variations of intensity $\delta I(t,x,W)$ and velocities $\delta
V(t,x,W)$ are mainly caused by oscillatory and convective motions.
Since here we are primary interested in convection, we filtered
out oscillatory motions in the Fourier space based on the
$k-\omega$ diagram \citep{Khomenko+Kostik+Shchukina2001,
Kostik+Khomenko2007, Kostik+etal2009}.

\subsection{Separation into granules and intergranular lanes}

We fragmented all spatial points into granular and intergranular points using
as a criterion their continuum intensity in \BaII\ line. We removed from the
consideration the dark pores (located mostly in intergranular lanes).
Intergranular bright points, characteristic for facular areas at high spatial
resolution (e.g. at 0.\arcsec1), are not distinguishable at the resolution of
our observations (see Fig.~\ref{fig:fov}). The second right panel of this
figure gives an example of the applied granular mask.

\section{Results}

To study the relationship between the magnetic field, velocity and intensity
in our data we proceeded in the following way. The magnetic field values,
measured in all pixels and in all five maps, were sorted by their magnitude,
and then divided into 20 groups (or bins), each of them containing an equal
number of pixels. Each bin was assigned a value of the magnetic field
strength by averaging the $B$ values over all points it contains. The \BaII\
intensity and velocity values (at different levels of the profile) were
measured and averaged separately for each group.
In addition to that, we have performed a similar procedure, but separately
for granular and intergranular points, previously segmented from all data.

\begin{figure}
\center
\includegraphics[width=9cm]{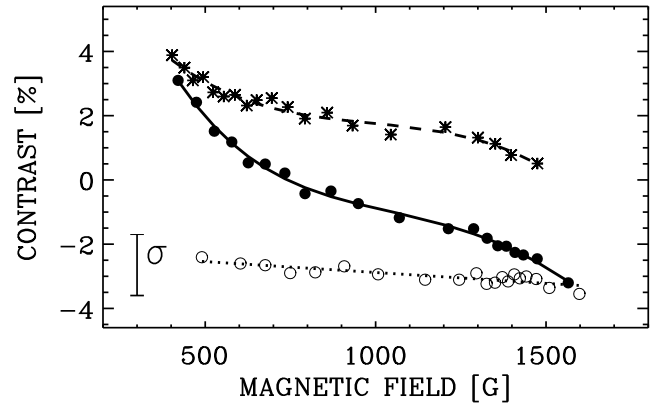}
\caption{Convective contrast measured from the \BaII\ line at the deepest atmospheric layer as
a function of magnetic field strength from \FeI\ lines. Black dots: all pixels
are considered; stars: only granular points; open circles: only intergranular
points.}
\label{fig:contrast}
\end{figure}

\begin{figure*}
\centering
\includegraphics[width=16cm]{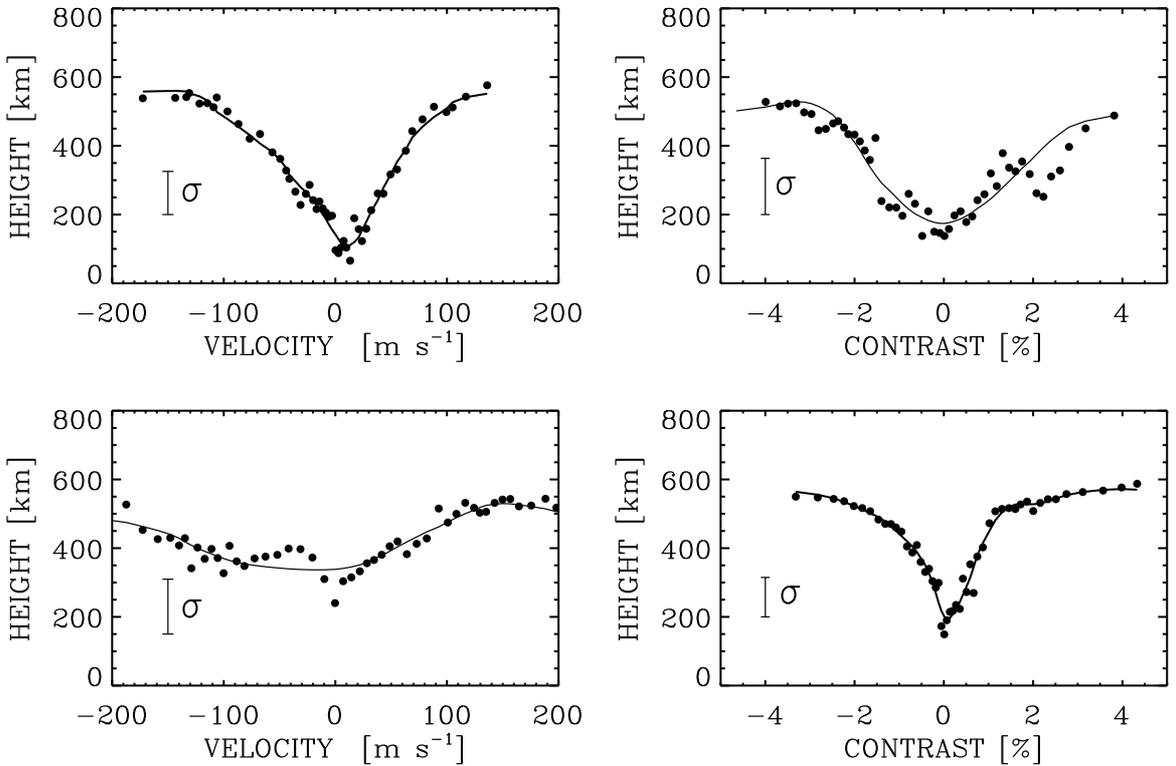}
\caption{Upper panels: height of the velocity sign reversal as a function of
``continuum'' velocity (left) and ``continuum'' contrast (right) from the
\BaII\ profile. The mean value is 330$\pm 160$ km. Each symbol is an average
over a bin with an equal number of data points. Bottom panels: height of the
contrast sign reversal as a function of the ``continuum'' velocity (left) and
``continuum'' contrast (right) from the \BaII\ profile. The average value is
420$\pm 130$ km. } \label{fig:reversal}
\end{figure*}

Figure \ref{fig:velocity} (left panel) shows values of the {\it unsigned}
convective velocities observed in \BaII\ line at different heights in the
solar atmosphere, binned for intervals of varying magnetic field strength.
The dependence on the magnetic field strength is not monotonic. The velocity
initially decreases with increasing magnetic field strength, reaching a
minimum near B = 1000 G, and then increases again.
A similar behaviour of the velocities is observed at all of 14 levels of the
\BaII\ line profile. For a more clear view, the figure only presents data for
three levels, corresponding, approximately, to heights of 0, 400, and
650 km. Note that unsigned velocities are represented and, thus, such a
behaviour is not surprising. It can be easily understood in terms of
magneto-convection, taking as example the behaviour of the quiet Sun
internetwork magnetic fields \citep[e.g.,][ see their Figure
7]{Khomenko+etal2003}. Indeed, as our observations are taken close to the
disk centre, the unsigned (absolute) values of the vertical velocities are
expected to be larger in granules (with weaker field) and in intergranular
lanes (generally containing strong magnetic field concentrations), being
small at the transition between granules and lanes (where the field is of
intermediate strength). The existence of regions with almost zero component
of the vertical velocity at the transitions between granules and lanes was
recently reported from the quiet Sun data taken by IMaX/{\sc Sunrise}
instrument \citep[][]{MartinezPillet2011}, see \citet{Khomenko+etal2010}.

The middle and right panels of Fig.~\ref{fig:velocity} show the behaviour of
the \BaII\ velocities separately for granules and lanes, this time only for
the deepest ``continuum'' atmospheric layer and taking into account their
sign. As can be seen, upward velocities of granules decrease with increasing
magnetic field strength, and the downward velocities in intergranular lanes
are increasing, in agreement with what was said above.

Figure~\ref{fig:contrast} shows the contrast variation measured at the
deepest ``continuum'' atmospheric layer from the \BaII\ profile binned for
magnetic field intervals of varying strength. The contrast is defined as a
standard deviation of the fluctuations of the quantity $\delta C$, see
Equation (\ref{eq:deltac}).
Here we compare three different statistics: (1) all pixels are considered
(black dots); (2) only granular points are considered (stars); and (3) only
intergranular points are considered (open circles). The contrast is given in
per cents with respect to the mean value over the whole observed area and
time, so that dark areas have negative contrast and bright areas have
positive contrast. In general, in all the cases, the contrast decreases with
magnetic field strength. When considering all the points, the contrast
changes from positive to negative as the magnetic field strength increases.
The reason is similar to the case of velocities, i.e. brighter granules have
generally weaker field than darker intergranular lanes, same as for the
magnetic fields in the quiet Sun. At the same time, it is interesting that
the dependence on contrast becomes much weaker if we consider statistics
separately for granules and lanes. For lanes, the dependence is within the
error bar limit, meaning that the contrast of intergranular lanes is nearly
independent of the field strength. We will discuss the implications of this
finding later in Section~\ref{sect:discussion}.


Another known feature of solar granulation in the quiet non-magnetic areas is
its contrast sign reversal and the velocity direction reversal observed at
the middle photosphere. Here we investigate this feature, but for the facular
area, searching for dependences of the heights of reversal on the strength of
granular motions at the continuum formation level, similar to
\citet{Kostik+etal2009}.

In our data, only about 37\% of elements show the contrast reversal with
height, and 26\% show the reversal of the direction of motion, as calculated
from the \BaII\ $\lambda$-meter velocities and intensities. We only used
these elements for the results presented in  Fig.~\ref{fig:reversal}.
The scatter of heights of the velocity and contrast sign reversals is very
high. These heights vary in a range from 50 to 650 km, i.e. spanning almost
the whole formation height range of the \BaII\ line profile.  Nevertheless,
if these heights are plotted as a function of ``continuum'' velocity or
``continuum'' contrast, there is a clear dependence, similar to the case of
the quiet Sun \BaII\ line data used in \citet{Kostik+etal2009}.
Fig.~\ref{fig:reversal} shows this result. We calculate the error bars as a
standard deviation over the values for each bin, and then average over all
bins.
The larger is the continuum velocity and contrast of convective elements, the
larger is the height where they change their direction of motion (upper two
panels). On average, this height is about 330 $\pm$ 160 km. Same happens for
the contrast sign reversal (bottom two panels).  The dependence on the
continuum contrast is more pronounced than on the velocity. On average, the
contrast sign reversal of the convective elements takes place at about 420
$\pm$ 130 km. Interestingly, both average reversal heights are about 100 km
larger than in the quiet Sun data of \citet{Kostik+etal2009} ($\approx$210
and $\approx$330 km, compared to 330 and 420 km).

Both the statistics over the quiet Sun data of \citet{Kostik+etal2009}, and
the facular data in the present study suggest that the velocity sign reversal
happens at lower heights that the contrast sign reversal. This should not be
interpreted in a way that the same ascending element first changes the
direction of motion at one height $h_1$ and then the contrast at another
height $h_2$, where $h_1<h_2$. We do the statistics for the contrast and
velocity sign reversals independently. Thus, we can have independent
elements, one changing the direction of motion at $h_1$ and another one
changing the contrast at $h_2$. Also, the descending elements can first
change their contrast and then the direction of motion at a lower height. We
do not differentiate between the descending and ascending elements, neither
between the bright and dark elements when calculating the reversal heights.
Note also, that only about small percentage of elements reverse their
contrast and direction of motion in our data.

\begin{figure}
\centering
\includegraphics[width=9cm]{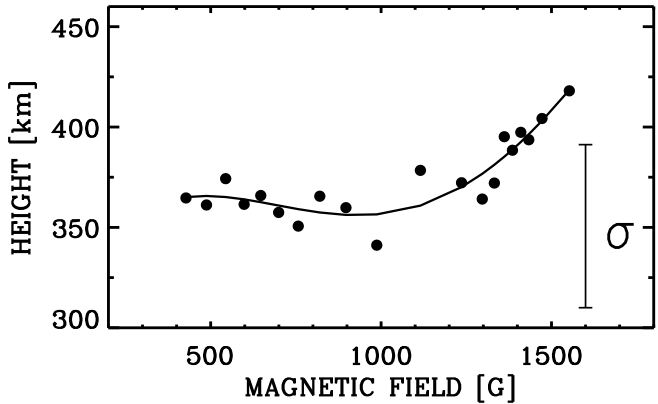}
\caption{Heights of the velocity and contrast sign reversal as a function of
magnetic field strength.} \label{fig:reversal2}
\end{figure}

Our next step was to investigate if the reversal height depends on the
magnetic field strength. To increase the statistics, we averaged
indistinguishably velocity and contrast reversal heights in all the data
points for bins of varying magnetic field strength. The result is given in
Figure~\ref{fig:reversal2}. As can be seen from the figure, there is a hint
for such a dependence within the error bar limit. The reversal height is
roughly independent of the magnetic field strength till about 1000 G. For the
larger field strengths, the reversal height increases from about 350 to 430
km.

\begin{figure}
\centering
\includegraphics[width=9cm]{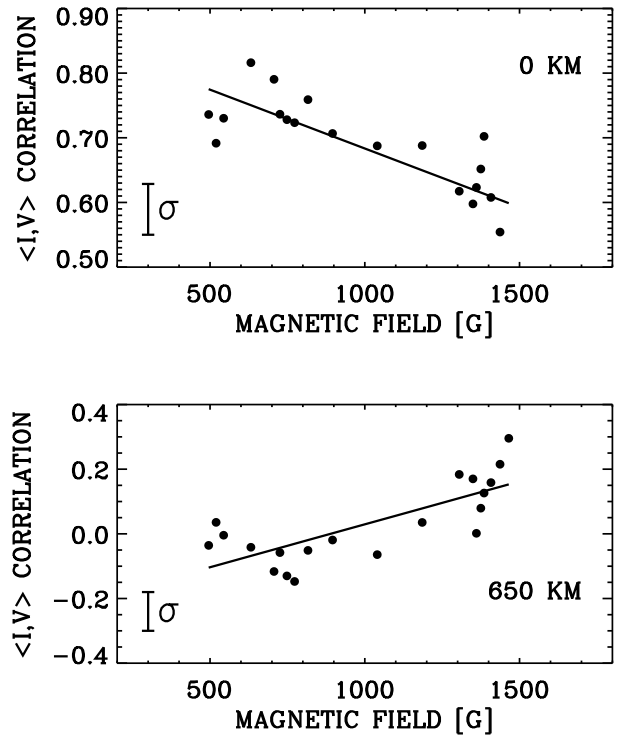}
\caption{Top: correlation coefficient between the ``continuum'' intensity and
``continuum'' velocity measured from the \BaII\ line profiles as a
function of the magnetic field strength. Bottom: same for the velocities and
intensities at the highest level $\sim$650 km.} \label{fig:correlation}
\end{figure}

Yet another important parameter characterizing granulation is the correlation
between intensity (contrast) and velocity of granular elements. This
correlation is expected to be maximum at the continuum formation layer,
decreasing to zero in the middle photosphere \citep[see e.g., ][ and
references therein]{Kostik+Shchukina2004}. The values of the correlation
measured by the latter authors do not exceed 75\% even in the deepest layer,
meaning that total coincidence of the intensity and velocity maxima is rare
for the solar photosphere. Here we investigate how the correlation
coefficient between the velocity and intensity changes as a function of the
magnetic field. Figure ~\ref{fig:correlation} shows this dependence for the
two atmospheric levels: the deepest ``continuum'' levels (top) and the
highest level around the \BaII\ line core ($\sim$650 km, bottom). It appears
that at the ``continuum'' level, the correlation is maximum for the
intermediate values of the magnetic field strength (500--700 G), reaching as
high as 77\%. The correlation coefficient decreases down to 60\% as the
magnetic field gets stronger. The variations between the maximum and minimum
values are larger than the statistical error bar (also shown in the Figure).
Unlike that, the correlation at 650 km is negative for intermediate magnetic
field strength, characteristic for inverse granulation. For larger field
strengths above 1000 G, the correlation turns positive reaching 20\% at
maximum. This result indicates that strong magnetic field in facular areas
help preserving the structure of convective elements till larger heights.

\begin{figure}
\centering
\includegraphics[width=9cm]{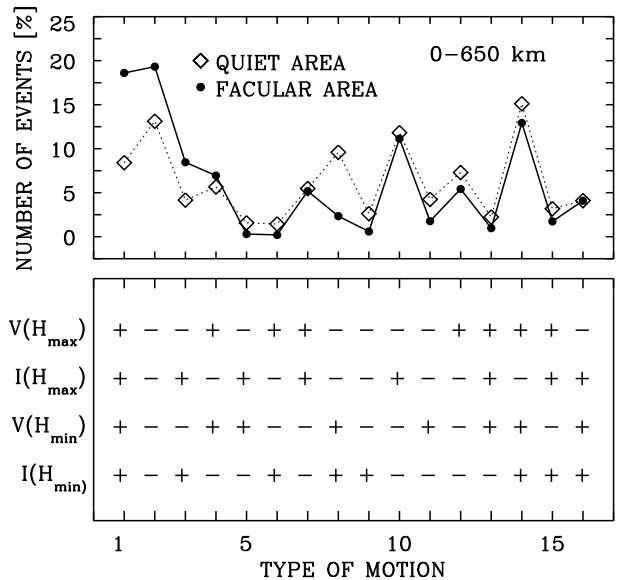}
\caption{Convective motions in the solar atmosphere according to
16-column model. Upper panel shows the number of cases corresponding
to each of the 16 types of motions (as defined at the bottom panel)
between 0 and 650 km. } \label{fig:16column}
\end{figure}

\section{Discussion}
\label{sect:discussion}

Before discussing our results, let us recall the medium spatial resolution
($\sim$0.\arcsec5  in the blue part of the spectrum) of our observational
data which, in principle, could affect our results.
In particular, we do not ``see'' manifestations of strong kG magnetic field
concentrations expected to be appear as bright points (BPs) in intergranular
lanes \citep{Title1992, Berger2004, Narayan+etal2010}.
Thus, there might be a chance that after segmenting all data points into
granular and intergranular points, unresolved BPs may appear belonging to
either as granular or intergranular groups, depending on the atmospheric
blurring.
Possibly, the second maximum (B = 1300 G) on the histogram of the magnetic
field strength distribution over granules (Fig.~\ref{fig:histogram}, middle
panel) is due to unresolved such BPs. For the purpose of visual comparison
the granular pixels with magnetic field above 1.2 kG are marked red in
Fig.~\ref{fig:fov}. At least visually, there is nothing particular about the
location of such pixels.
To check if the unresolved BPs affect our results, we have repeated all the
relevant calculations presented above, but this time excluding from
consideration the granular points with magnetic field strength above 1.1 kG
(a value corresponding to a local minimum on the histogram in
Fig.~\ref{fig:histogram}, middle panel). We have not found any significant
differences in our results.

The dependence of convective velocities on magnetic field strength can be
compared to the results of similar studies by other authors.
Measurements for the plage and facular areas are not that frequent and the
most adequate comparison can be done with the works of \citet{Title1992,
Montagne1996, Morinaga2008, Narayan+etal2010}. These authors used different
method for obtaining magnetic field strength from their measurements
(inversion of spectropolarimetric data or magnetogram signals). Also, data of
different spatial resolution (between 0.\arcsec15 and 1.\arcsec0) and at
different wavelengths (spectral lines from 4305 \AA\ to 6768 \AA) are
presented. Thus direct one-to-one comparison with our results is not
possible. Nevertheless, we can state that there is a qualitative agreement
between our results and those from the works mentioned above, as for velocity
distributions. The granular motions in plage areas on average exhibit upflows
for intermediate field strength below 600 G. For the larger fields above 600
G the structures show tendency for downflows \citep[except for ][ where
downflows are dominant for the magnetogram signal above 100 G]{Title1992}.
Our results presented in Fig.~\ref{fig:velocity} better agree with the
results by \citet{Morinaga2008}, though the latter authors study the velocity
flows in the vicinity of a solar pore,  unlike the quiet plage analyzed here.

\citet{Evans+Catalano1972}, and \citet{Holweger+Kneer1989}, were apparently
the first ones to discover the contrast sign reversal with height in the
solar atmosphere.
Later, this phenomenon was investigated by \citet{Kucera+Rybak+Wohl1995,
Espagnet+Muller+Roudier+Mein+Mein1995, Kostik+Shchukina2004,
Puschmann+etal2005, Stodilka2006, Janssen+Gauzzi2006}.
\citet{Cheung+etal2007} have shown from numerical MHD simulations, that the
temperature (and, thus contrast) reversal in the photosphere is due to the
competition between expansion and cooling of rising fluid elements and their
radiative heating. The heights of contrast reversal derived from observations
by different authors vary in a wide range from 60 to 350 km.
\citet{Kostik+etal2009} pointed out that this scatter may be attributed to
the difference in the formation heights of the spectral lines used in
different observational studies.
More recently, it was found that not only contrast reverses sign in the solar
atmosphere, but also the direction of motion of convective elements reverses
with height \citep{Kostik+Shchukina2004, Kostik+etal2009}. These studies were
performed for granulation observed in very quiet non-magnetic regions.
Our results from Fig.~\ref{fig:reversal} point out that both contrast and
velocity sign reversal take place in the facular regions as well. The
moderate magnetic field in facula (400--900 G), does not affect the height
where the reversal takes place compared to the quiet solar regions
(Fig.~\ref{fig:reversal2}). However, stronger field above 1 kG ``forces'' the
convective elements to travel over larger vertical distances without changing
the sign of contrast and direction of motion.
Apparently, strong magnetic field in facular areas promotes more effective
energy transfer in the upper layers of the solar atmosphere by convective
elements, since they can reach larger heights.
At the first instance, it might seem that this conclusion is in disagreement
with the results shown in the top panel of Figure \ref{fig:correlation}
regarding the correlation between velocity and contrast. The correlation
coefficient almost linearly decreases with increasing magnetic field.
However, the correlation at the heights of formation of the \BaII\ line core
behaves in the opposite way (bottom panel of Figure \ref{fig:correlation}).
While the absolute values of the correlation in the upper atmosphere are
lower, as expected, the correlation is, on average, positive for larger
magnetic field strengths above 1 kG, indicating that the structure of
convective elements is maintained till that height.

Yet another prove of the latter result is provided by
Figure~\ref{fig:16column}. This figure compares the types of motion observed
in the quiet area \citep[data from][]{Kostik+etal2009} and in our facular
area. The bottom panel of the figure indicates our classification of the
types of motion: sign ``+'' indicates convective elements moving upwards or
those whose contrast is above the average; sign ``-'' indicated downward
moving elements or those with the contrast below the average. We compare the
velocity sign and contrast of convective elements at the ``continuum'' level
and the at highest level available for us around the formation of the \BaII\
line core. Interestingly, we find that the number of convective elements that
do not change neither contrast nor velocity direction in the facular area to
be about 53\% (motions of type 1 to 4). Compared to this, in the quiet area
only 31\% of convective elements behave this way. We conclude that the strong
magnetic field of the facular region helps to stabilize the convective
elements and allow the convective energy to reach larger heights.

Possibly the most interesting result of our study is the one presented in
Fig.~\ref{fig:contrast}. The general decrease of the contrast with increasing
magnetic field (if all the points are considered indistinguishably) seems to
be a natural result and confirms the results by the earlier studies
\citep{Title1992, Topka+etal1992, Montagne1996, Topka+etal1997, Berger2007,
Kobel2011}. However, the behaviour of intergranular pixels, considered
separately from the rest, is significantly different (dashed line in Figure
\ref{fig:contrast}). Namely, the intergranular contrast is essentially
independent of the magnetic field strength within the error bar limit.
The same conclusion was reached earlier by \citet{Berger2007}. The authors
claimed that in facula regions close to the limb one rather detects granular
walls, that should be essentially independent of the magnetic field strength.
Here we confirm this finding but for the facular region close to the disk
centre.

The decrease of the contrast with increasing magnetic field obtained in
previous studies seems to be a result of the particular distribution of the
magnetic field strength in these structures (Fig.~\ref{fig:histogram}). The
granules have generally weaker fields, while intergranules have stronger
ones. By averaging indistinguishably the intensity values for different $B$
intervals, one would naturally obtain a larger number of granular pixels for
the bins with lower B (i.e., larger contrast); and the opposite for
intergranules. Thus, it is important to segment granular and intergranular
pixels prior to analyzing the dependence of the contrast on $B$.

All the statistical dependences found in our paper are more significant that
the corresponding error bars, given in all the figures. The only exception is
the dependence shown in Fig.~\ref{fig:reversal2}, which is within the error
bar. The error bars were calculated as standard deviation over the values
averaged for a given bin.

The advantage of our work over the similar studies is the real magnetic field
measurements obtained from the inversion of extremely magnetically sensitive
\FeI\ spectra in the infrared, instead of using a magnetogram signal.
The alignment of our data sets from both instruments was done with a 1 pixel
precision (0.\arcsec185), improving the statistical significance of our
results.
It is important to recall, however, that the diffraction limits in the
visible and in the infrared are very different, leading to a naturally lower
spatial resolution of the infrared data. For a better alignment we
artificially reduced the spatial resolution of the continuum \BaII\ maps to
make it comparable with the one of TIP Stokes $I$ maps.
The lower resolution of TIP data might, in principle affect the precise value
of the magnetic field derived from this data, since the magnetic field
patches are more diffused and magnetic elements of mixed polarities might
cancel. However, the cancelations are expected to be more important in the
mixed-polarity inter-network areas, and not in a (mostly unipolar) plage area
as the observed one. Since we derive the magnetic field from the inversion of
Stokes profiles, using a stray-light component as a free parameter, this
improves our accuracy in the magnetic field strength determination.
The magnetic field values given at the horizontal axis of
Figs.~\ref{fig:velocity}, \ref{fig:contrast},  \ref{fig:reversal2}, and
\ref{fig:correlation} are averages over the intensity bins. While these
average values might, in principle, be affected by the cancelations in a
mixed-polarity area, we do not expect this to be important in a uni-polar
plage.
Another possible source of error is that our data do not have enough
resolution to resolve intergranular bright points. However, as pointed out at
the beginning of this section, we have cross-checked that removing the
``granular'' points with $B>1.2$ kG from our analysis (as possible unresolved
bright points) provides the same results, in particular the one given in
Fig.~\ref{fig:contrast}. This confirms once more our conclusion that the
contrast of intergranular lanes is independent on the magnetic field, in
agreement with \citet{Berger2007}.

It will be interesting in the future to check the results of our work using
data with a larger spatial resolution and maintaining the advantages of the
simultaneous Stokes polarimetry and velocity/intensity measurements till the
upper photosphere.

\section{Conclusions}

Using simultaneous spectropolarimetric observations in the infrared \FeI\
lines at 1.56 $\mu$m and spectral observations of the strong
line of \BaII\ 4554 \AA\, formed high in the atmosphere,
with a moderate temporal resolution, we have
investigated the behaviour of the convective component of the intensity and
velocity fields in the facular region close to the solar disk centr. The
\BaII\ line allowed us to perform a study over a wide range of heights, from
the continuum formation height up to about 650 km in the solar atmosphere.  We
have reached the following conclusions:

\begin{enumerate}

\item At the bottom photosphere, the convective velocities of granules
    decrease with magnetic field strength, while the convective
    velocities of intergranules increase with magnetic field strength.

\item Similar to the quiet regions, we detect that in facular regions the
    contrast of granulation and the velocities of convective elements
    reverse their sign with height. The height where such reversal takes
    place depends on the strength of convective elements at the bottom
    photosphere, but also on the magnetic field strength. Convective
    elements above stronger magnetic features reach higher in the
    atmosphere without breaking.

\item The correlation coefficient between velocity and intensity at the
    bottom-most level decreases with the magnetic field. In the upper
    atmosphere, above 500 km, the correlation gets larger with increasing
    the magnetic field.

\item The contrast of convective elements decreases with increasing
    magnetic field strength. But if intergranular lanes are considered
    separately, their contrast is nearly independent of the field
    strength.

\end{enumerate}

\begin{acknowledgements}
Based on observations made with the VTT telescope operated on the island of
Tenerife by the Kiepenheuer-Institut fur Sonnenphysik in the Spanish
Observatorio del Teide of the Instituto de Astrofisica de Canarias.  This
research has been supported by the Spanish Ministry of Economy and
Competitiveness (MINECO) under the grants AYA2010-18029 and AYA2011-24808.
The authors are grateful to Manuel Collados for the help with observations
and useful discussions.
\end{acknowledgements}


\end{document}